\documentclass{article}
\usepackage{graphicx}
\usepackage[utf8]{inputenc}
\usepackage{amsmath}
\usepackage{titling}
\usepackage{amsfonts}
\usepackage{amssymb}
\usepackage[margin=2.54cm]{geometry}
\usepackage{parskip}
\usepackage[super,square,numbers,sort&compress]{natbib}
\usepackage{hyperref}

\newcommand{\subtitle}[1]{%
  \posttitle{%
    \par\end{center}
    \begin{center}\Large#1\end{center}
   }%
}

\newcommand{\subsubtitle}[1]{%
  \preauthor{%
    \begin{center}
    \large #1 \vskip0.5em
    \begin{tabular}[t]{c}
    }%
}

\title{Literature Review}
\subtitle{Implementation of Formal Semantics and the Potential of Non-Classical Logic Systems for the Enhancement of Access Control Models}
\author{Alvin Tang}
\date{22 May 2023}

\begin{document}

\maketitle

\section{Introduction}

Access control is a significant component of computer systems to safeguard the security of information by limiting the actions and operations that a user can perform. As an important tool to manage permissions of genuine users and processes to retrieve and manipulate data\cite{principles}, system architects must ensure that the access control schemes and models work feasibly in the desired manner efficiently. To provide a rigorous way for verifying the correctness of the theories, formal logic is extensively utilised to describe access control rules and policies. Given the importance of rigorousness ad comprehensiveness in this field of study, it is worthwhile to explore the features of formal semantics and the underlying classical logic principles upon which most existing research outputs are established. In our proposed research, we will discover how formal logic is applied in well-known access control models, how the features of classical logic lead to limitations of the existent theories, as well as how non-classical logic systems may contribute to the enhancement thereof.

We shall hereby review the existing discoveries by computer scientists by comparing the characteristics and purposes of different access control models, analysing a few sets of theories formalising access control schemes and the most significant limitations in existing access control models. In the meantime, we will briefly explore the constraints of classical logic and the notable features of alternative logic systems with literature by a few computing researchers and logicians. These revelations will help us find out the importance of improving existing models, the potential applications of non-classical logic for access control and the aspects of which we have to be mindful during the research.

\section{Current Knowledge}

\subsection{Significance of Access Control for Cyber Security}

The purpose of access control is to limit the range of files and programs to which a genuine user may gain access for reading, writing and execution. Various security components of a computer system serve distinct purposes to shield the integrity and safety of information by providing multiple layers of protection for programs and data. There is no exception for access control as its implementations are generally integrated with other cyber security tools and protocols. Their policies cannot be enforced without the use of a reference monitor, which mediates every attempted access to resources in a system.\cite{principles} Complete mediation is achieved when architects of computer systems designate the reference monitor as the sole point of access control decision-making so that there is no way a legitimate or intruding user/process may bypass the policies.

In the meantime, authentication of user identities is critical to ensure that all individuals and processes have no alternative way to access the system without abiding by the permission rules. Whilst the development of access control models has significance for discovering ways to organise and manage permissions, research on hashing, password policies and multi-factor authentication likewise contribute to the efficient implementation of access control policies by preventing malicious users from being able to conduct fraudulent activities with imposture of others' identities. Results of our research project, consequently, should integrate well with other layers of security protection in computer systems.

The implementation of access control is ubiquitous in everyday computer systems. Linux, for example, implements access control at multiple levels based on the principle of least privilege. In addition to firewall rules and recording of file ownership details, Linux uses permission bits to determine whether a user may access a file or directory in line with the categorisation of their identities.

The permission bits in Linux correspond to the user, the user's group and others. Each category of users may be assigned permission to read, write and execute a file.\cite{access-control-assessment} Take \texttt{/etc/shadow} as an example. This system file documents the list of users registered in the system and the hash of their passwords. In most Linux installations, the file has permissions and ownership which resembles the following.

\begin{verbatim}
    [alice@mysystem ~]$ ls -l /etc/shadow
    -rw------- 1 root root 900 Mar 25 16:42 /etc/shadow
\end{verbatim}

The \texttt{rw} indicates that the owner (\texttt{root}) has read and write permissions. The subsequent dashes implies that other users have no permission to read or modify the file.

This access control list (ACL) design provides us with the assertion that an ordinary user or anyone that has access to the computer system will not be in the position to view the password hashes of other users, which could be maliciously used for impersonation. With a well-implemented reference monitor, the purpose of access control to mitigate the risks can be achieved as a result.

In this example with the \texttt{/etc/shadow} file, the password hashes should be kept protected as the use of obsolete hashing algorithms (\textit{e.g.} MD5) may enable the exploitation of login credentials with preimage or collision attacks. Even if modern hashing functions (\textit{e.g.} SHA-512) which are popular in systems we use nowadays are currently considered safe, advancing technology and progress from the constant scrutiny of these algorithms by mathematicians and computer scientists may render them otherwise. Access control acts as a layer of protection supplementary to hashing and, therefore, provides us with stronger confidence in the security of the systems.

In summary, the significance of access control for enhancing cyber security involves authentication of user identity, authorisation to access data and monitoring of user activities to enforce permission policies and comply with cyber security regulations and standards.\cite{principles} Linux, as one of the most mainstream operating systems especially for servers, demonstrates the importance of access control to achieve the principles of cyber security. Researchers have praised the ease of implementation with ACL. However, its inability to specify access rights for individual users and specific groups\cite{access-control-assessment} has motivated us to delve into alternative access control models as there are more sophisticated requirements for complex environments.

\subsection{Different Access Control Models for Various Purposes}

In addition to ACL, researchers have developed a vast assortment of access control models for varying levels of security and fulfilment of security requirements. Albeit the discretionary access control (DAC) system implemented by Linux is simple to manage, it might not provide adequate granularity and flexibility. The reliance of DAC on the delegation of privileges to users or groups may be concerning in some settings, such as in the military or intelligence organisations, when unintended authorisations to access certain information may lead to detrimental effects. To resolve the challenges, more sophisticated models such as the mandatory access model (MAC), the role-based access model (RBAC) and the attribute-based access model (ABAC) are utilised.

With mandatory access control, security policies are completely controlled by the security administrator. MAC is varied from DAC in that users do not have control over resources or determine who has access thereto.\cite{models} With this design, we may deduce that operations such as \texttt{chmod} or \texttt{chown} by ordinary users would not be allowed if MAC was applied in typical Linux systems.

Role-based access control is often preferred by large entities due to its greater adaptability and scalability. As opposed to DAC and MAC, where the access policy is defined for each file separately, RBAC is designed around the idea that users are assigned roles within an organisation.\cite{models} As an illustration, suppose there are two users \texttt{alice} and \texttt{bob} in an organisation such that they are assigned the roles as follows.

\begin{center}
\begin{tabular}{c|c}
    User & Roles\\
    \hline
    \texttt{alice} & \texttt{student}, \texttt{COMP2100-student}, \texttt{COMP2550-student}\\
    \texttt{bob} & \texttt{student}, \texttt{COMP2300-student}, \texttt{COMP2550-student}\\
    \texttt{charlie} & \texttt{staff}
\end{tabular}
\end{center}

Let the following files be assigned with their corresponding roles as follows.

\begin{center}
\begin{tabular}{c|c}
    File & Roles\\
    \hline
    \texttt{student\_info.html} & \texttt{student}, \texttt{staff}\\
    \texttt{comp2100\_assignment.txt} & \texttt{COMP2100-student}\\
    \texttt{comp2300\_lecture.mp4} & \texttt{COMP2300-student}\\
    \texttt{research\_proposal.pdf} & \texttt{COMP2550-student}
\end{tabular}
\end{center}

In this scenario, \texttt{alice} has access to \texttt{comp2100\_assignment.txt} whilst \texttt{bob} does not. On the contrary, \texttt{comp2300\_lecture.mp4} is exclusively accessible by \texttt{bob}. Meanwhile, both of them have permission to read \texttt{research\_proposal.pdf}. The webpage \texttt{student\_info.html} is available for all three individuals.

In RBAC, both users and files may be assigned multiple roles. Furthermore, the assignment of roles to the latter may specify the extent of permissions. For instance, read and write permissions may be assigned to the \texttt{staff} role for \texttt{student\_info.html} while read-only permission may be assigned for the \texttt{student} role.

RBAC is advantageous when the scale of a system is sizable where it is strenuous and time-consuming to manage permission roles for every file individually. With roles apparently indicating the rights and responsibilities of each user, this model also thoroughly applies the separation of duties principle\cite{principles} by only giving access to resources they need to perform.

Although the clarity of role assignment for users makes RBAC easily understandable and systematised, the level of control provided by this model may not be adequate for dynamic environments. As a solution, decisions are made in attribute-based access control grounded in the evaluation of attributes associated with the user or resource.\cite{models} As a demonstration, suppose \texttt{alice}, \texttt{bob}, \texttt{charlie} and \texttt{dennis} are students in a university. Let \texttt{department} and \texttt{year} be attributes indicating the department to which they belong along with the year in which they are studying.

\begin{center}
\begin{tabular}{c|cc}
    User & \texttt{department} attribute & \texttt{year} attribute\\
    \hline
    \texttt{alice} & \textit{Computing} & 1\\
    \texttt{bob} & \textit{Social Sciences} & 1\\
    \texttt{charlie} & \textit{Computing} & 2\\
    \texttt{dennis} & \textit{Physics} & 3\\
\end{tabular}
\end{center}

Suppose there are multiple files stored in the university database with the following access policies.

\begin{center}
\begin{tabular}{c|c}
    File & Access policy\\
    \hline
    \texttt{computing\_notes.pdf} & Allow access $\iff$ \texttt{department} is \textit{Computing}\\
    \texttt{maths\_notes.pdf} & Deny access $\iff$ \texttt{department} is \textit{Social Sciences}\\
    \texttt{new\_students.html} & Allow access $\iff$ \texttt{year} equals 1\\
    \texttt{returning\_students.html} & Allow access $\iff$ \texttt{year} is greater than 1
\end{tabular}
\end{center}

As a result, the access permissions to the files by the four users are as follows.

\begin{center}
\begin{tabular}{c|cccc}
    Access granted? & \texttt{alice} & \texttt{bob} & \texttt{charlie} & \texttt{dennis}\\
    \hline
    \texttt{computing\_notes.pdf} & Yes & No & Yes & No\\
    \texttt{maths\_notes.pdf} & Yes & No & Yes & Yes\\
    \texttt{new\_students.html} & Yes & Yes & No & No\\
    \texttt{returning\_students.html} & No & No & Yes & Yes
\end{tabular}
\end{center}

While roles in RBAC are discrete and enumerable, attributes in ABAC offer more flexibility as they can be, for instance, numerical values such as the \texttt{year} attribute above. Access policies in ABAC are customisable for more intricate situations and dynamic environments. In the above example, assume that all students proceed to a higher year every two semesters. Compared to the RBAC where we have to reassign the roles for students every year, we just have to increment the \texttt{year} attribute for all students by one in ABAC. In other words, ABAC outperforms alternatives in terms of flexibility to define access policies with conditions and efficiency to modify users' permissions by altering discrete or continuous attributes. It is suitable in situations where the attributes of entities may be constantly changing, such as in cloud computing or e-commerce platforms.

In summary, there are various access control models designed by researchers to meet different security requirements and solve challenges in varying circumstances. DAC and MAC are advantageous for their simplicity whilst RBAC and ABAC offer more granular control over access policies with their sophisticated design. The successive step of analysis is to discover how we can mathematically and logically represent these models and the advantages of doing so.

\subsection{Use of Formal Logic in Access Control Models}

Computer science researchers have been using formal logic and semantics to discover rigorous frameworks for specifying access control policies to prove the correctness of models, to analyse the implications of policies and to structure proposals of new theories. RBAC, being one of the most favoured access control models in the industry, are frequently scrutinised by academics for verification, vulnerabilities and improvements.

Logic has been the foundation for the formal verification of the RBAC model. For example, members of the IEEE Computer Society proposed using set theory and sequents to not only conduct security analysis but also to explore the computation complexity of the model implementation. In the research, it is suggested that in reality configurations are often modified as resources may start being shared at some point and stop being accessible sometime later.\cite{formal-verification} Taking the dynamic environment into consideration, there is a need to ensure that permission authorisations and the revoking thereof should be free from inadvertent effects.

In order to conduct a security analysis, we first have to define an access control scheme formally. With reference to this research, it is defined as the tuple $\langle \Gamma, Q, \vdash, \mathcal{A}, \Sigma, \Psi \rangle$\cite{formal-verification} where
\begin{itemize}
    \item $\Gamma$ is a set of states,
    \item $Q$ is a set of queries,
    \item $\vdash: \Gamma \times Q \to \{ \textit{true}, \textit{false} \}$ is the function determining whether a query is true or not in a state,
    \item $\mathcal{A}$ is a set of principals,
    \item $\Sigma$ is a set of actions, and
    \item $\Psi$ is a set of state-transition rules.
\end{itemize}

We understand that in this context the researchers are viewing access control schemes as state-transition systems, similar to finite-state automata except start or terminal states are not explicitly defined. Using this definition, it is evident that states correspond to information available for making access control decisions whereas a query refers to an access request.

Let $\gamma \in \Gamma$ and $q \in Q$. Then, the function $\vdash$ is understood as follows.
\begin{itemize}
    \item The statement $\gamma \vdash q$ means approval of the request.
    \item The statement $\gamma \nvdash q$ equals denial thereof.
\end{itemize}

We may apply this state-transition model of verification to our RBAC system example in Section 2.1. We can define the users as the set of principals $\mathcal{A} = \{ \texttt{alice}, \texttt{bob}, \texttt{charlie} \}$. The question of whether \texttt{alice} has access to \texttt{student\_info.html} is thus a query.

Solely with our RBAC setting in Section 2.1, we are incapable of defining the relationships or hierarchies between varying roles. It is not sensible to assign an individual with the \texttt{COMP2100-student} role but not the \texttt{student} role though nothing is preventing us from doing so. The logic relation framework advocated by this state-transition model introduces relations to form a hierarchy of roles.

According to their definition, $RH \subseteq \mathcal{R} \times \mathcal{R}$, where $\mathcal{R}$ is the set of roles, is an irreflexive acyclic relation.\cite{formal-verification} Suppose $r_{1}, r_{2} \in \mathcal{R}$. Then, $r_{1} \succeq_{RH} r_{2}$ is equivalent to the statement ``every user who is authorised for $r_{1}$ is also authorised for $r_{2}$''.

Back to our RBAC example. We may define $RH$ such that
$$\forall r \in \mathcal{R}_{\texttt{student}} . r \succeq_{RH} \texttt{student}$$
\begin{center}
    where
\end{center}
$$\mathcal{R}_{\texttt{student}} = \{ \texttt{COMP2100-student}, \texttt{COMP2300-student}, \texttt{COMP2550-student} \}.$$

The roles assigned to \texttt{alice} and \texttt{bob} may then be simplified without affecting the actual policy:
\begin{center}
\begin{tabular}{c|c}
    User & Roles\\
    \hline
    \texttt{alice} & \texttt{COMP2100-student}, \texttt{COMP2550-student}\\
    \texttt{bob} & \texttt{COMP2300-student}, \texttt{COMP2550-student}\\
    \texttt{charlie} & \texttt{staff}
\end{tabular}
\end{center}

Until now we have not yet taken into account changes in roles. Bearing the above state-transition model definition in mind, we may consider the assignment and revocation of roles as actions in the set $\Sigma$. For example, $\textit{assign}(u_{a}, u_{t}, r_{t})$ refers to the assignment of the role $r_{t}$ for the user $u_{t}$ by the user $u_{a}$.

In reality, such actions should only be done by some particular users (\textit{e.g.} system administrators). To verify whether the action is credible (\textit{i.e.} whether the user $u_{a}$ has the authority to perform the assignment), we check the roles possessed by $u_{a}$ and their implications indicated in the state against the action.

Using the state-transition model, the set of actions\cite{formal-verification} is defined as $$\chi = \{ \textit{assign}(u_{a}, u_{t}, r_{t})\ |\ u_{a}, u_{t} \in \mathcal{U} \land r_{t} \in \mathcal{R} \} \cup \{ \textit{revoke}(u_{a}, u_{t}, r_{t})\ |\ u_{a}, u_{t} \in \mathcal{U} \land r_{t} \in \mathcal{R} \}.$$

In our situation,
\begin{itemize}
    \item $\mathcal{U} = \{ \texttt{alice}, \texttt{bob}, \texttt{charlie} \}$, and
    \item $\mathcal{R} = \{ \texttt{staff}, \texttt{student}, \texttt{COMP2100-student}, \texttt{COMP2300-student}, \texttt{COMP2550-student} \}$.
\end{itemize}

Quoting the researchers' definition, an revocation in $\chi$ succeeds if and only if
\begin{enumerate}
    \item the user $r_{t}$ is already assigned with the role $r_{t}$, and
    \item the assigner has the authority to revoke the role from a user fulfilling the conditions.
\end{enumerate}
These three requirements are further formalised by relations and tuples. For simplicity, we can imagine that \texttt{charlie} is the course convenor for COMP2550. It has the authority to assign and revoke the \texttt{COMP2550-student} role fulfilling the conditions. Since a student cannot enrol in both COMP2550 and COMP4550 simultaneously, \texttt{comp2550-student} and \texttt{comp4550-student} are mutually exclusive.

For instance, suppose \texttt{bob} finds the coursework too demanding and wish to drop one of the courses. We would like to specify that only \texttt{charlie}, the staff member, has the authority to modify the authorisations.

To prove that $\textit{revoke}(\texttt{charlie}, \texttt{bob}, \texttt{comp2550-student})$ is valid, we may logically show that
\begin{enumerate}
    \item \texttt{bob} possess the \texttt{comp2550-student} role.
    \item \texttt{charlie} has the authority to revoke \texttt{comp2550-student} from an individual.
\end{enumerate}

There are lots of considerations we may take into account, such as the implications of assignment or revocation of roles, as well as how we may optimise the efficiency of the RBAC models with role hierarchies by normalising role assignments. The analysis of RBAC systems with this state-transition model is only one of the wide variety of methods proposed by researchers. The abundant research into access control models demonstrates the value of formal logic and semantics in formal verification, security analysis and more. In our upcoming research, we will perform critical analyses comparing different ways of logical formalisation for RBAC schemes. We shall discover the types of logic systems used, the considerations taken into account in various research efforts and to what extent their theories most neatly describe and facilitate a granular control of permission management.

\subsection{Development of Alternative Logic Systems}

Classical (orthodox) logic dominates in the field of mathematics and scientific research yet its limitations, such as the material paradoxes arising from its definitions, motivate scholars to develop alternative logic systems. Some prominent substitutions are namely fuzzy logic, relevant logic and intuitionistic logic.\cite{alternative-logic}

Classical logic is built on Boolean algebra, where truth values are binary. A statement is either \textit{true} (1) or \textit{false} (0). It is onerous to reason for uncertain information or vague statements in this conventional logic system, leading to the sorites paradox.\cite{alternative-logic} For instance, suppose a thousand grains of sand is a \textit{heap} of sand whilst one single grain is not. It is difficult to justify whether some amount of sand is considered a \textit{heap} when the number of grains is between one and a thousand. Fuzzy logic extends the truth values into a continuous spectrum between 0 (absolute falsity) and 1 (definite truth). A truth value of 0.8, for instance, indicates a higher level of truth than 0.2.

Relevant logic resolves the material paradox from the principle of explosion caused by the definition of implication in orthodox logic. In classical logic, the statement ``\textit{p implies q}'' is defined as follows:
\begin{center}
\begin{tabular}{cc|c}
    $p$ & $q$ & $p \to q$\\
    \hline
    0 & 0 & 1\\
    0 & 1 & 1\\
    1 & 0 & 0\\
    1 & 1 & 1
\end{tabular}
\end{center}
When the premise $p$ is false, the entire implication statement is true. Logicians universally agree upon this definition with the justification that a statement is false iff a contradiction exists. In fact, there is no requirement for $p$ and $q$ to be interrelated. As a consequence, the statement ``if $1 + 1 = 3$, the Sun is a planet'' is a totally legitimate theorem in orthodox logic. Relevant logic is constructed upon axioms so that it requires premises and conclusions in implication statements to be inextricably linked. As a result, some proof techniques such as vacuous discharge in natural deduction are no longer valid in relevant logic.\cite{relevant-logic}

The principle of excluded middle is a noteworthy characteristic of classical logic. The double negation of a statement is always equivalent to itself ($\neg \neg p \equiv p$). This might not always accurately transform our statements in natural language or common sense into rigorous logical notations. For example, the statement ``it is not that I am not happy'' is usually not regarded as interchangeable with ``I am happy''. Intuitionistic logic disallows double negation elimination in logic proofs, precluding the principle of excluded middle.\cite{alternative-logic}

By understanding the features of alternative logic systems, we can critically analyse how their characteristics can be utilised for applications in the field of cyber security and, in particular, creating bettered access control schemes. We shall refer to the logic theory knowledge in our third research question (Section 4.3).

\section{Present Lack of Knowledge and Research Gap}

In spite of the multitude of access control models available, each model has its restraints. Some of the incapabilities are attributed to the characteristics of classical logic and the dependency of research on this traditional logic system. Our inquiry aims to address this void in the existing research.

\subsection{Limitations in Existing Access Control Models}

Existing access control models such as DAC, MAC, RBAC and ABAC have often been implemented successfully in the industry. There are nonetheless drawbacks in these models. With RBAC, it may be difficult to handle situations where multiple roles are required to perform certain actions, or where roles have to be dynamically assigned based on contextual information. Notwithstanding the flexibility of ABAC to handle dynamic environments effectively, it might be hard to handle conflicts between attribute values.

To discuss the most substantial research gaps in this field of study, we can first refer to the definition of limitations in an access control model summarised by academics as the failings of one or more of the following characteristics: \cite{limitations}
\begin{itemize}
    \item Inescapable: inability to break security policies by circumventing access controls set by the model.
    \item Invisible: seamless user and administrative interaction with the model.
    \item Feasible: cost-efficiency and practicality to implement the model.
\end{itemize}

In the following sections, we shall analyse the logic models offered by researchers for RBAC and ABAC, and analyse how classical logic may result in not being able to fulfil the above requirements.

\subsubsection{Inescapability: Comprehensiveness of Logical Establishment and Verification}

In a study conducted by Fatima \textit{et al.} to identify the disadvantages of existing RBAC models, several extensions to RBAC models are developed to suit context-sensitive and dynamic environments.\cite{rbac-vs-abac} However, they are still inadequate to address the remaining limitations to offer fine-grained, content-based, and multi-factored solutions.

From our analysis, we observe that these unresolved limitations are to a considerable extent caused by the shortcomings and characteristics of classical logic. In Section 2.3, we mentioned that in RBAC, some roles can be defined as incompatible with others. In complex organisations or extensive systems, there may be an immense number of roles to be managed. In the real-world application of access control or during the design of the access control policies, system engineers might not be capable of detect role conflicts immediately. Meanwhile, most research depends on classical logic to perform formal verification to demonstrate the correctness and rigorousness of these models. When principals in RBAC have contradictory roles, we might be able to deduce anything in accordance with the principle of explosion and, hence, struggle to rule out mistakes efficiently when the undetected conflicts are taken as premises of logic statements. This results in the concern that the security policies are not inescapable, inadvertently leaving risks caused by security vulnerabilities.\cite{rbac-vs-abac}

We may refer to the publication \textit{A Modal Logic for Role-Based Access Control} by Kosiyatrakul \textit{et al.}, one of our references in the research proposal about the use of modal logic, an extension to classical logic, to represent role inheritance and role-permission associations with partial orders and sequents. In their research, modal logic is adopted to facilitate representations of statements about necessity and possibility.\cite{modal-logic}

For example, the situation when the reference monitor makes an approval decision for the statement ``user $U$ , acting in authorised role $R$, makes a request $q$'' is logically represented with the statement
$$(U\ \texttt{for}_{RA}\ R\ \texttt{says}\ q) \supset q$$

where
\begin{itemize}
    \item $R$ is a role,
    \item $U \in \texttt{authorised\_roles}(R)$, and
    \item $q \in \texttt{authorised\_permissions}(R)$.
\end{itemize}

The author utilises the concepts ``necessity'' and ``'possibility'' in modal logic to establish the role hierarchies for rigorous definitions of the functions $\texttt{authorised\_roles}$ and $\texttt{authorised\_permissions}$. Although their proposed proof system efficiently achieves separation of duty, it is not well-designed for the resolution of conflicts, such as circular references in role hierarchies. Modal logic does not handle the situation when a role inherits from another and \textit{vice versa} concurrently.

In our planned research, we should explore the different ways to resolve these limitations with various approaches, such as allowing vagueness in the determination of access control decisions and requiring relationships between premises and conclusions in implication statements to avoid role explosion.

\subsubsection{Invisibility and Complexity: Ease, Effectiveness and Practicality of Implementations}

The research, in conjunction with the comparison between RBAC and ABAC, suggests that the driving force behind various existing ABAC-based solutions is the same basic idea revolving around the use of attributes of the subject, object, and environment for making access control decisions.\cite{rbac-vs-abac} They suggest a standardisation for the theoretical foundations of ABAC to establish a more powerful and theoretical footing. In order to achieve this, we should never disregard the potential disadvantages when we extend the access control system and develop logic models for them.

In ABAC, attributes can be discrete or continuous values.\cite{abac-logic} Restrictions on access to videos on online streaming platforms based on users' geographical locations are examples of using a discrete attribute (country/region) for access control. This model is more advisable for precise control over security policies than most other models but in spite of the flexibility to integrate continuous values as attributes (inputs), it cannot handle uncertainty or ambiguity in its decisions (outputs). Architects of the scheme have to precisely define the boundaries of the attributes. Being potentially arduous to justify or reach a consensus on the boundaries, the burden to manage ABAC may no longer be negligible.

The model examples we have considered so far have a shared characteristic of having binary truth values for access control decisions. Referring to the roles of a user, which are also binary in the sense that a user either has a role or otherwise, the system must decide definitely whether a user is granted or denied access to some resource. As an extension to the RBAC example we have in Section 2.2, we may formulate an access control policy to deny access to a resource iff a user has the \texttt{student} role (statement \#1). If we are convinced that an individual is a \texttt{student} if and only if not being a \texttt{staff} member, designers of an access control system may assume that this policy is identical to allowing access to the resource iff the user has the \texttt{staff} role (statement \#2).

In real-world scenarios, this is often not the case. Even if we have rigorous mechanisms to disallow a user from possessing both the \texttt{staff} and \texttt{student} roles concomitantly by defining them as mutually exclusive, we still have not defined the situation when an individual is neither a \texttt{student} nor a \texttt{staff} member. By the principles of classical logic, the entity would be allowed access according to statement \#1 but denied access with reference to statement \#2. A paradox arises. The definitions for mutual exclusivity of roles will require a tedious fundamental restructuring when we expand the number of roles in a system, which raises concerns for the feasibility of existing models in some situations.

\subsection{Applying Non-Classical Logic for Enhanced Access Control Models}

Alternative logic systems, such as fuzzy logic, relevant logic and intuitionistic logic, have been applied in several fields of computer science research. However, despite the great potentiality to utilise non-classical logic systems to enhance access control models, there is a paucity of research regarding this topic.

Fuzzy logic is a technique to develop artificial intelligence. Given that this unconventional logic system does not require numerical precision, the modelling machining process in AI can be tuned up efficiently according to varying control conditions. As a result, the resultant AI models may respond quickly to convoluted sensory inputs.\cite{fuzzy-logic-ai} To our understanding, estimation with probability is a major component of machine learning and thence fuzzy logic can help with the creation of such models.

Relevant logic and intuitionistic logic have been applied in the analysis and affirmation of programming language semantics. Logic programming, although not the most applied programming paradigm in the industry, is a robust tool for data management and natural language processing. Analysts have used these two alternative logic systems to prove the equivalence of logic programs, taking advantage of the fact that intuitionistic logic cannot be described by a finite set of truth values.\cite{logic-programs}

The existing research effort to apply alternative logic systems clearly demonstrates that they truly contribute to the field of computing. Nevertheless, there is very narrow research in academia regarding the application of alternative logic systems in cyber security. Principally, there is a limited amount of research on the complexity analysis and quantification of the efficiency of access control models. Notwithstanding that alternative logic systems have seldom been used for access control models or other information security technologies, it can be helpful to investigate the benefits of their application and to fuel the motivation for realising the conceptual models into deployable systems.

The limitations of existing access control models discussed in Section 3.1 become concerning as we encounter prevailing challenges in the security, usability and efficiency of the models. In order to advance research in this domain, we will explore the application of these three non-classical logic systems in access control, drawing parallels to how AI employs them for estimation and logic programming incorporates non-binary truth values.

\section{Research Questions and Aims}

Taking into account our knowledge of access control models and the promising prospect for their enhancement to benefit information security, the proposal for the research is summarised with three preeminent questions to be put under examination.

\subsection{Formal Logic in Existing Access Control Models}

Our first aim of the proposed research is to analyse the use of formal logic in existing access control models. On top of the state-transition model for RBAC analysed in Section 2.2, we will discover a few other logic proof systems and modelling theories proposed by other researchers. From the research papers we reviewed above, we recognise that academics promote the evolution of logic models catering to different needs for access control.

By conducting in-depth analyses of existing logic establishment and verification methods, we can find out the importance of formal logic to confirm that the access control systems are inescapable, invisible and feasible. When we recommend enhancements to the models with non-classical logic, we shall be attentive to assure that these requirements are not violated.

\subsection{Classical Logic Features to Limitations of Existing Models}

The second major research question is to investigate how features of classical logic result in limitations of the existing models. In connection with the flaws in existing access control models mentioned in Section 3.1, it is observed that existent models may be unable to tackle the prevalent challenges in dynamic or complicated environments.

Whilst researchers have scrutinised existing models a lot and applied extensions to classical logic (\textit{e.g.} modal logic, temporal logic), they do not tackle the most elemental weaknesses of this long-established logic system. We shall extend our investigation in Section 3 to conduct the exploration.

\subsection{Non-Classical Logic for Enhanced Models}

Our final goal of discovery is to explore the feasibility of applying non-classical logic to form a foundation for revising access control models which may address the issues with existing ones. The ultimate goal of this research is to not only establish logical ways to describe and verify access control models that tackle limitations caused by orthodox logic but also to justify that the new models are practicable. Hence, the methodologies, including simulations of models with Haskell programs, mentioned in the research proposal help us certify that the models are viable and appropriate for real-world applications.

\section{Conclusion}

The aforementioned findings highlight the crucial role of access control in cyber security. Given the multifarious collection of access control implementations available, careful consideration must be given to selecting appropriate models based on the specific requirements of the application environment. Formal logic has been an integral aspect of the research and development process. Throughout our investigation, we maintain a constant focus on the completeness of the logic models and ensure that any proposed enhancements adhere to the principles of well-designed models.

\newpage

\bibliographystyle{IEEEtranN}
\bibliography{references.bib}

\section*{Acknowledgements}

The generative AI results by ChatGPT contributed to the refinement of language in this literature review. It also provided tips regarding \LaTeX\ syntax and the format of citations.

\end{document}